\documentclass[epj]{svjour}
\usepackage[utf8]{inputenc}
\usepackage{graphics}
\usepackage{graphicx}
\usepackage[T1]{fontenc} 
\usepackage{multirow} 
\usepackage{float} 
\begin{document}
\title{Anisotropic Lifshitz holography in Einstein-Proca theory with stable negative mass spectrum}
\author{Roberto Cartas-Fuentevilla\inst{1}\thanks{{e-mail: rcartas@ifuap.buap.mx}}\and Alfredo Herrera-Aguilar\inst{1}\thanks{{e-mail: aherrera@ifuap.buap.mx}}\and V. Matlalcuatzi-Zamora\inst{1}\thanks{{e-mail: vmatlal@ifuap.buap.mx}} \and Uriel Noriega\inst{2}\thanks{{e-mail: 201128576@alumnos.fcfm.buap.mx}}\and \\Juan M. Romero\inst{3}
\thanks{{e-mail: jromero@correo.cua.uam.mx}}%
}                     
%
%
\institute{Instituto de F\'isica, Ben\'emerita Universidad Aut\'onoma de Puebla, Apdo. Postal J-48, Puebla, Puebla 72570, M\'exico\and Facultad de Ciencias F\'isico Matem\'aticas, Ben\'emerita Universidad Aut\'onoma de Puebla, Apdo. Postal 165, Puebla,\\Puebla 72000, M\'exico\and Departamento de Matem\'aticas Aplicadas y Sistemas, Universidad Aut\'onoma Metropolitana-Cuajimalpa, M\'exico,\\CDMX 05300, M\'exico}
\date{Received: date / Revised version: date}
%
\abstract{
In this article we focus on constructing a new family of spatially anisotropic Lifshitz spacetimes with arbitrary dynamical exponent $z$ and constant negative curvature in $d + 1$ dimensions within the framework of the Einstein-Proca theory with a \textit{single} vector field. So far this kind of anisotropic spaces have been constructed with the aid of a set of vector fields. We also consider the spatially isotropic case as a particular limit. The constructed metric tensor depends on the spacetime dimensionality, the critical exponent and the Lifshitz radius, while the curvature scalar depends just on the number of dimensions. We also obtain a novel spectrum with negative squared mass, we compute the corresponding Breitenlohner-Freedman (BF) bound and observe that the found family of spatially anisotropic Lifshitz spaces respects this bound. Hence these new solutions are stable and can be useful within the gravity/condensed matter theory holographic duality, since the spectrum with negative squared mass is complementary to the positive ones already known in the literature. We also examine the null energy condition (NEC) and show that it is essentially satisfied along all the boundary directions, i.e. along all directions, except the $r$ one, of our Lifshitz spacetime with the corresponding consistency conditions imposed on the scaling exponents.
} 
\maketitle
\section{Introduction}
\label{intro}
The main idea of the gauge/gravity duality is that we can establish a relationship between the physics of a field theory in $ d $ dimensions and the physics of gravity in $ d + 1 $ dimensions. Gauge/gravity duality is of fundamental importance since it provides new links between quantum theory and gravity. It also has led to new concepts in mathematics and physics, and provides new tools for solving problems in many areas of theoretical physics \cite{Taylor,Taylor1,Ammon}. 

The most important and so far best understood example of this holographic principle is the so-called anti-de Sitter/conformal field theory (AdS/CFT) correspondence, proposed by Maldacena \cite{maldacena} (for a modern comprehensive review see \cite{nastase,20}). This duality establishes a relationship between backgrounds with negative curvature and conformal field theories with one less (non-compact) spatial dimension. 

One of the more interesting extensions of this duality consists in the study of quantum critical systems that belong to the realm of condensed matter physics\cite{Hartnoll,Herzog,McGreevy,Hartnoll2,Horowitz,HLS,Horowitz2,ZQHH,FosterLiu}. These quantum critical models are scale invariant where space and time scale in a different way \cite{Hertz}.  On the other hand, these quantum critical theories are difficult to approach by means of perturbative methods based on weakly interacting quasiparticles. Moreover, these systems exhibit non-relativistic scaling symmetries of the following form
 \begin{eqnarray}\label{lif}
 t \longrightarrow t'=\lambda^{z}t, & \quad  & x^{i}\longrightarrow x^{'i}=\lambda x^{i}
\end{eqnarray}
where $\lambda$ is an arbitrary real constant with dynamical critical exponent $z\ne 1$. This kind of symmetries is realized at quantum critical points and provides a strong kinematic relation to different extensions of the AdS/CFT correspondence. The Lifshitz scaling transformations (\ref{lif}) are present in the asymptotic symmetry group of the gravitational dual theory.
\\

It is worth mentioning that since 1941 Lifshitz showed that systems that are invariant under the anisotropic scaling transformation (\ref{lif}) appear within the framework of condensed matter physics \cite{L}.  Notably, field theories invariant  under the anisotropic scaling transformations (\ref{lif}) improve their  high energy behavior \cite{anselmi,Farias,Hatanaka,Alexandre,Alexandre2,Horava0,Chen,Alexandre3,Alexandre4}.  In particular, using these  transformations,   Ho\v{r}ava formulated a modified gravity  which seems to be ghosts free and power counting renormalizable  for $z=3$ \cite{Horava}.  This modified   gravity has dynamical inconsistencies \cite{Henneaux}, but  quite soon healthy extensions of it  were found \cite{Pujolas,pujolas2}. Remarkably,  Ho\v{r}ava gravity predicts that the spectral dimension of spacetime reduces at  short distances \cite{Horavad}. Other interesting properties of the Ho\v{r}ava gravity can be found in \cite{horava1,horava2,horava3,horava4,horava5,horava6,horava7,horava8,horava9}.

In order to construct metrics that obey the anisotropic Lifshitz scalings  (\ref{lif}), an energy-momentum tensor with the same symmetries is required to source the gravitational field. A first realization of this idea was based on four-dimensional gravity coupled to a vector field and a 2-form that interact topologically \cite{KLM}. However, an equivalent and simpler formulation was obtained within the framework of gravity coupled to a massive \textit{single} vector field \cite{Taylor,CHK}. It turns out that this additional structure beyond pure Einstein gravity realizes Lifshitz geometries as ground states. On the other hand, it is important to understand the physics that is encoded in the gravitational theory when the Proca vector field is included, since the latter is hodge dual to the 2-form action originally considered in \cite{KLM}.  An important aspect that it is worth mentioning is that even when observables within gravitational systems can be difficult to characterize, due to the dynamical nature of spacetime, these observables can be defined on the boundary of such spacetimes (as in the case of Lifshitz geometries) \cite{HLS}.

Within the context of the holographic duality, a bulk field corresponds to an operator in the dual field theory. The essential holographic dictionary was first formulated by Gubser-Klebanov-Polyakov and Witten (GKPW) \cite{GKP,Witten}. Besides, in \cite{HLS} we have at hand a comprehensive review of the connection between bulk fields and operators in the dual field theory that turns out to be completely general. There we observe that the important relation is
\\
\\
\begin{tabular*}{\columnwidth}{@{\extracolsep{\fill}}llll@{}}
\centering
Field theory source $h$&$\Longleftrightarrow$ & Leading behavior $\phi_{(0)}$ of bulk field       &   $\ast$        \\
\end{tabular*}
\\
\\
\begin{tabular*}{\columnwidth}{@{\extracolsep{\fill}}llll@{}}
\centering
Field theory expectation value $< O >$ &$\Longleftrightarrow$ & Subleading behavior  $\phi_{(1)}$ of bulk field      &  $\ast\ast$   \\
\end{tabular*}
\\
\\
Here $\phi_{(0)}$ is the boundary value $h$ of $\phi$ (where $\phi$ is a bulk field), and $\phi_{(1)}$ is the expectation value $<O>$ of $\phi$. Furthermore, it is important to mention that since the relations ($\ast$) and ($\ast\ast$) are completely general, then, as concrete examples, the sources could be the chemical potential or an electric field and the expectation values are the charge density and the electric current respectively \cite{HLS}.

Motivated by the fact that Lifshitz spacetimes are currently relevant since they are duals to quantum critical theories used to describe strongly coupled systems in condensed matter. Concrete examples of such systems are high temperature superconductors, whose physical properties are governed by non-relativistic quantum critical points, and unitary fermions, i. e. fermion whose interaction are fine-tuned to produce a non-relativistic scale invariant system. Moreover, non-relativistic holography is interesting by itself as an example of a gauge/gravity correspondence without asymptotically AdS spacetimes, and, hence it could shed some light on fundamental and general questions of the holographic principle \cite{Taylor}. Therefore, in this paper we shall focus on constructing a new family of exact solutions to the Einstein-Proca equations that represent spatially anisotropic Lifshitz spacetimes, with arbitrary dynamical exponent $z$ and constant negative curvature in $d + 1$ dimensions. 

This paper is organized as follows. In section 2 we review Anti-de Sitter and Lifshitz spacetimes. In section 3 we introduce the Einstein-Proca theory with a \textit{single} vector field and quote a new family of exact solutions to the Einstein-Proca field equations that have spatially anisotropic Lifshitz symmetry. We continue (in subsection 3.1) with a discussion on the novel spectrum with negative squared mass that respects a Breitenlohner-Freedman bound and comment on the null energy condition in subsection 3.2. In subsection 3.3 we consider a spatially isotropic Lifshitz background as a particular limit, and we find the BF bound for this spacetime. We conclude in section 4 with a short summary and conclusions.
\section{Anti-de Sitter and Lifshitz spacetimes}
 \vskip 5mm
 \textbf{Anti-de Sitter spacetimes}
 \vskip 5mm
The geometry of the Anti-de Sitter spacetime is given by the following metric (in Poincar$\grave{e}$ coordinates)
\begin{equation}
ds^{2} = l^{2}\left[r^{2}\left(-dt^{2} +  dx^{2}_{i} \right)+ \frac{dr^{2}}{r^2}\right]\,, 
\end{equation}
where $i=1,2,...,D$ in a spacetime of $D+2$ dimensions, $l$ is the AdS radius \cite{nastase}. The boundary of AdS is found in $r\rightarrow\infty$. Anti-de Sitter space is a solution of the Einstein equations with a negative cosmological constant $\Lambda=-\frac{D(D+1)}{2}$ described by the action
\begin{equation} \label{accion ads}
S = \int d^{d+1}x \sqrt{-g} \left[ R +D(D+1) \right] \,,
\end{equation}
where $d$ is the number of spatial dimensions of this gravity theory and  $d=D+1$; then Einstein equations are
\vskip 3mm
\begin{equation}\label{eq:eh}
R_{\mu\nu}-\frac{1}{2}g_{\mu\nu}R-\frac{D(D+1)}{2} g_{\mu\nu}=0,
\end{equation}
\\
and  the curvature scalar reads
\begin{equation}
R=-\frac{(D+1)(D+2)}{l^2} =-\frac{d(d+1)}{l^2} \,,
\end{equation}
\\
which evidently is negative definite for any dimension and possesses an arbitrary factor given by $l^{-2}$. 
\\
\\
\\
\textbf{Lifshitz spacetimes}
\vskip 3mm
On the other hand a Lifshitz invariant theory is spatially isotropic and homogeneous, i.e., is invariant under space and time translations and spatial rotations:  
\begin{eqnarray}\label{eqs}
H:&  & t \longrightarrow t'=t+a;\nonumber \\
P^{i}:&  & x^{i}\longrightarrow x^{'i}=x^{i}+a^{i}; \\
L^{ij}:&  & x^{i}\longrightarrow x^{'i}=L^{i}_{j}x^{j},\nonumber 
\end{eqnarray}
in addition, a Lifshitz theory admits the non-relativistic scaling symmetry
 \begin{eqnarray}\label{eq2}
D_{z}:&  & r \longrightarrow r'=\lambda^{\pm 1}r\nonumber \,, \\
&  & t \longrightarrow t'=\lambda^{\mp z}t, \\
&  & x^{i}\longrightarrow x^{'i}=\lambda^{\mp 1} x^{i} \,. \nonumber
\end{eqnarray}
where the parameter $z$ is the dynamical exponent. The symmetry group consisting of ($H$, $P^i$, $L^{ij}$, $D_{z}$) will be denoted as ${\bf{Lif}}_{D}(\bf{z})$. 
Furthermore, a generalization of a Lifshitz invariant theory takes place when considering a spatially anisotropic metric, i.e. when now a Lifshitz theory admits the following non-relativistic scaling symmetry
 \begin{eqnarray}\label{eq3}
D_{z}:&  & r \longrightarrow r'=\lambda^{\pm 1}r\nonumber \,, \\
&  & t \longrightarrow t'=\lambda^{\mp z}t, \\
&  & x^{i}\longrightarrow x^{'i}=\lambda^{\mp z_{i}} x^{i} \,, \nonumber
\end{eqnarray}
where the $z_{i}$ are critical exponents along different spatial directions.
The simplest Lifshitz geometry \cite{KLM} is given by the following metric
\begin{equation} 
ds^{2} = \ell^{2} \left( -r^{\pm 2z} dt^{2} + r^{\pm 2} dx^{2}_{i} + \frac{dr^{2}}{r^{2}}\right),  \label{metric}
\end{equation}
where  $0\leq r<\infty$, $\ell$ is the Lifshitz radius and again $i = 1, 2, ..., D$ in a spacetime of $D + 2$ dimensions. This metric admits the space and time translations and spatial rotations (\ref{eqs}) and the scaling symmetry (\ref{eq2}). The '$+$' and '$-$'  signs are related via the coordinate transformation $r\rightarrow \frac{1}{r}$ which leaves the metric invariant. Although, the matter fields of the considered theory can have different behavior under the transformation $r\rightarrow \frac{1}{r}$, since only the field invariants preserve its form under this inversion.

\section{Einstein-Proca theory and Lifshitz spacetimes}

In this section, we study the Einstein-Proca theory of a \textit{single} massive vector field coupled to gravity with a negative cosmological constant. Thus,  we use the following action\footnote{It is important to note that there is no gauge invariance of the action (\ref{eqbulkactiong}).}
\begin{equation}\label{eqbulkactiong}
S = \int d^{d+1}x \sqrt{-g} \left[R + D(D +1) - \frac{1}{4} F^2 - \frac{1}{2} M^2 A^2 \right], \
\end{equation}
where $F_{\mu\nu}$ is the vector field strength,  $A_{\mu}$ is a massive vector field and $M$ is its mass, and the Greek indices run from $0$ to $D+1$. The Einstein-Proca equations of motion following from this action are
\begin{equation}\label{eq:Eing}
R_{\mu \nu} - \frac{1}{2} g_{\mu\nu} R - \frac{D(D+1)}{2} g_{\mu\nu}
= \frac{1}{2} \left( F_{\mu \rho} F_{\nu}{}^\rho -\frac{1}{4} g_{\mu \nu} F^2\right)
+ \frac{M^2}{2} \left( A_\mu A_\nu -\frac{1}{2} g_{\mu\nu} A^2 \right),\
\end{equation}
whereas the Proca equations read
\begin{equation}\label{eq:Procag}
\nabla_\mu F^{\mu \nu} - M^2 A^\nu = 0 \,.
\end{equation}
In this paper we will be interested in obtaining new spatially anisotropic Lifshitz spacetimes of the Einstein-Proca theory.
Thus, we shall start with the following metric ansatz
\begin{equation}
\label{eqmetricag}
ds^{2} = \ell^{2} \left( -f(r) dt^{2} + P_{i}(r) dx^{2}_{i} + \frac{dr^{2}}{r^{2}}\right),
\end{equation}
where $f(r)$ and $P_i(r)$ are arbitrary functions of the extra coordinate and  again $i = 1, 2, ..., D$. Spatially anisotropic black branes with Lifshitz scaling were perturbatively constructed in \cite{Dibakar}. Without loss of generality we shall consider the case in which the only non-trivial component of the vector field is $A_{t}$. This requirement is consistent with the fact that the mixed components of the energy-momentum tensor $T_{tr}$, $T_{ti}$ and $T_{ir}$ must vanish since we are considering a diagonal metric ansatz. Therefore under the ansatz (\ref{eqmetricag}) the Einstein-Proca field equations consist of the following nonlinear ODEs for the unknown functions $f(r)$, $P_{i}(r)$ and $A_{t}(r)$:
\begin{eqnarray} \label{eq tt 2g}
-\frac{r^{2}P_{i}}{2}\frac{P''_{k}}{P_{k}}\!+\!\frac{3r^{2}P_{i}}{8}\frac{P_{k}^{'2}}{P_{k}^{2}}\!-\!\frac{r^{2}P_{i}}{8}\frac{P'_{l}P'_{k}}{P_{l}P_{k}}\!-\!\frac{rP_{i}}{2}\frac{P'_{k}}{P_{k}}\!+\!\frac{D(D+1)}{2}\ell^{2}P_{i}\!=\!\frac{r^{2}P_{i}}{4\ell^{2}f}A_{t}^{'2}\!+\!\frac{M^{2}P_{i}}{4f}A_{t}^{2},
\end{eqnarray}
\begin{eqnarray} \label{eq rr 2g}
-\frac{r^{2}P_{i}}{8}\frac{P_{k}^{'2}}{P_{k}^{2}}+\frac{r^{2}P_{i}}{8}\frac{P'_{l}P'_{k}}{P_{l}P_{k}}+\frac{r^{2}P_{i}f'}{4f}\frac{P'_{k}}{P_{k}}-\frac{D(D+1)}{2}\ell^{2}P_{i}=-\frac{r^{2}P_{i}}{4\ell^{2}f}A_{t}^{'2}+\frac{M^{2}P_{i}}{4f}A_{t}^{2},
\end{eqnarray}
\begin{eqnarray} \label{eq xx 2g}
\nonumber -\frac{r^{2}P''_{i}}{2}\!-\!\frac{r^{2}f'P'_{i}}{4f}\!-\!\frac{rP'_{i}}{2}\!-\!\frac{r^{2}P'_{i}}{4}\frac{P'_{k}}{P_{k}}\!+\!\frac{r^{2}P^{'2}_{i}}{2P_{i}}\!+\!\frac{r^{2}P_{i}f''}{2f}\!+\!\frac{rP_{i}f'}{2f}\!-\!\frac{r^{2}P_{i}}{4}\left(\frac{f'}{f}\right)^{2}\!+\!\frac{r^{2}P_{i}}{2}\frac{P''_{k}}{P_{k}}
\!+\!\\ \frac{r^{2}P_{i}f'}{4f}\frac{P'_{k}}{P_{k}}\!+\!\frac{rP_{i}}{2}\frac{P'_{k}}{P_{k}}\!+\!\frac{r^{2}P_{i}}{8}\frac{P'_{l}P'_{k}}{P_{l}P_{k}}\!-\!\frac{3r^{2}P_{i}}{8}\frac{P_{k}^{'2}}{P^{2}_{k}}\!-\!\frac{D\!(\!D\!+\!1\!)}{2}\ell^{2}\!P_{i}\!=\!\frac{r^{2}P_{i}}{4\ell^{2}f}A_{t}^{'2}\!+\!\frac{M^{2}P_{i}}{4f}A_{t}^{2},
\end{eqnarray}
\begin{equation} \label{Procaeqng}
A''_{t}+\left(\frac{1}{2}\frac{P'_{k}}{P_{k}}-\frac{1}{2}\frac{f'}{f}+\frac{1}{r}\right)A'_{t}-\frac{M^{2}\ell^{2}}{r^{2}}A_{t}=0,
\end{equation}
where primes denote derivatives with respect to the $r$ coordinate and repeated indices $k$ and $l$ denote summation from $1$ to $D$.

It turns out that the curvature scalar (or Ricci scalar) for the metric ansatz (\ref{eqmetricag}) reads
\begin{eqnarray} \label{eqRg}
R= g^{\mu\nu}R_{\mu\nu}&=&\frac{r^{2}}{2\ell^{2}}\left[\!-\!2\frac{f''}{f}\!-\!2\frac{f'}{rf}\!+\!\frac{f^{'2}}{f}\!-\!\frac{f'}{f}\frac{P'_{k}}{P_{k}}\!-\!2\frac{P''_{k}}{P_{k}}\!+\!\frac{3}{2}\frac{P^{'2}_{k}}{P^{2}_{k}}\!-\!2\frac{P'_{k}}{rP_{k}}\!-\!\frac{1}{2}\frac{P'_{l}P'_{k}}{P_{l}P_{k}}\right].
\end{eqnarray}
In order to solve the field equations (\ref{eq tt 2g})-(\ref{Procaeqng}), it is convenient to use the two following {\it{ans$\ddot{a}$tze}}
\begin{equation}\label{eq:fg} 
f(r) = r^{\pm2z}, 
\end{equation}
\begin{equation} \label{p1g}
P_{i}(r) =  r^{a_{i}},
\end{equation}
where $z$ is an arbitrary number and the $a_{i}$ are arbitrary constants.  As a result of substituting these {\it{ans$\ddot{a}$tze}}
 into the field equations (\ref{eq tt 2g})-(\ref{eq xx 2g}), we obtain a system of algebraic equations for $a_{i}$ with the following independent solutions:
\begin{eqnarray}\label{a1}
a_{1}\!=\!\frac{1}{2}\left[\!\mp\!2z\!-\!\!\displaystyle\sum_{m=3}^{D}\!a_{m}\!\pm\!\sqrt{-2\!\displaystyle\sum_{m=3}^{D}\!a_{m}^{2}\!-\!\!\displaystyle\sum_{m,n=3}^{D}\!a_{m}a_{n}\!\mp\!4z\!\displaystyle\sum_{m=3}^{D}\!a_{m}\!-\!12z^{2}\!+\!8D(D\!+\!1)\ell^{2}}\right],
\end{eqnarray}
\begin{eqnarray}\label{a2}
a_{2}\!=\!\frac{1}{2}\left[\mp 2z\!-\!\!\displaystyle\sum_{m=3}^{D}\!a_{m}\!\mp\!\sqrt{\!-\!2\!\displaystyle\sum_{m=3}^{D}\!a_{m}^{2}\!-\!\!\displaystyle\sum_{m,n=3}^{D}\!a_{m}a_{n}\!\mp\!4z\!\displaystyle\sum_{m=3}^{D}\!a_{m}\!-\!12z^{2}\!+\!8D(D\!+\!1)\ell^{2}}\right],
\end{eqnarray}
and for arbitrary ${a_{3}, a_{4}, a_{5},...,a_{D} }$ subjected to the following restriction
\begin{equation}\label{eq:sum}
 \displaystyle\sum_{k=1}^{D}a_{k}=\mp2z,
\end{equation} 
which manifestly depends on the spacetime dimensionality $D$ and the critical exponent $z$, and allows for a plethora of powers of $r$ for the metric functions $P_{i}(r)$ different from those so far reported in the literature \cite{Taylor1}. 

Since all the $a_{m}$ are completely arbitrary, by choosing them to be positive, $a_{m}$$>$$0$, we shall have a negative definite $a_{2}$, while the sign of $a_{1}$ will depend on whether the sum of the first and second terms is smaller or greater than the third one; however their difference can be made positive by appropriately setting the curvature radius $\ell$. For instance, when $D=2$, if $\ell^{2}$$>$$\frac{z^{2}}{3}$, then $a_{1}$$>$$0$; when $D=3$, if $\ell^{2}$$>$$3z^{2}$, for $a_{3}\sim z$, then $a_{1}$$>$$0$. Thus, with the aid of $\ell^{2}$ we can have as much positive $a_{m}$ as we wish.
\\

For these metric functions the non-vanishing component of the massive vector field is 
\begin{equation} \label{kag}
A_t= c\, r^{\pm z}, \qquad \qquad  c=\frac{\sqrt{2D(D+1)}\ell^{2}}{z}.
\end{equation}
On the other hand, the parameters of the theory read
\begin{equation} \label{eqm2g}
M^2 = -\frac{z^2}{\ell^2},
\end{equation} 
\begin{equation}\label{eq:ac}
\displaystyle\sum_{k=1}^{D}a_{k}^{2}=4D(D+1)\ell^{2}-4z^{2}.
\end{equation}
An interesting and novel feature of the aforementioned family of Lifshitz solutions is that it possesses a spectrum with negative squared masses. However, these modes of the massive vector field must satisfy a Breitenlohner-Freedman bound for the system to be stable (see next Section).
\\
\\
Thus, the full metric is expressed as follows

\begin{equation}\label{eqmg}
ds^{2}=\ell^2\left(-r^{\pm2z}dt^{2}+\frac{1}{r^{2}}dr^{2}+r^{a_{i}} dx^{i}dx_{i}\right),
\end{equation}
which is invariant under the spatially anisotropic Lifshitz transformations (\ref{eq3}) if the coordinates transform in the following way:
 \begin{eqnarray}
& & r \longrightarrow r'=\lambda r,\nonumber \\
&  & t \longrightarrow t'=\lambda^{\mp z}t,  \\
&  & x^{i}\longrightarrow x^{'i}=\lambda^{-{\frac{a_{i}(z, D, \ell^2)}{2}}} x^{i} \, \label{eq:ansatz_e}.\nonumber
\end{eqnarray}
\vskip 3mm
The curvature scalar upon substitution of the metric functions $f$ and $P_{i}$ in (\ref{eqRg}), reads
\begin{equation}
R=-D(D+1);
\end{equation}
here we note that the curvature scalar only depends on the dimensionality $D$, it is constant, negative and independent on the critical exponent $z$. There are several works where the squared mass and the curvature scalar depend on both the dimensionality $D$ and the dynamical exponent $z$ as well (see for instance \cite{Taylor,Taylor1,LL}) within the same theory. It would be interesting to find new solutions and/or the conditions under which these quantities acquire dependence on both $D$ and $z$.  
\subsection{A Breitenlohner-Freedman bound for our Lifshitz field configuration}

It is well known that in AdS spacetime tachyons (particles with their negative mass square) can arise, causing an instability only if their squared mass falls below a negative value. The allowed range for the negative squared mass is obtained from calculating the Breitenlohner-Freedman bound, guaranteeing the energy positivity of the system and, hence, its stability \cite{BF}.
Thus, in spite of having a negative spectrum for $M^{2}$, given by (\ref{eqm2g}), these values can be allowed in spacetimes with negative curvature if these satisfies the Breitenlohner-Freedman bound,  that renders positive values for the energy of the system.
\vskip 4mm
In particular, in order to find the Breitenlohner-Freedman bound for our case, we shall consider the spacetime given by the background (\ref{eqmg}) and the Proca equations without gravitational back-reaction, i.e.,  in the perturbative limit in which the massive vector field  $A(t, r, x_{i})$ does not alter the structure of spacetime. The Proca equations
\begin{equation}\label{eqProca1g}
\nabla_\mu F^{\mu \nu} - M^2 A^\nu = 0
\end{equation}
lead to the following constraint  in curved spacetime
\begin{equation}\label{eq:lorentz}
\nabla_{\nu}A^{\nu}=\partial_{t}A^{t}+\partial_{r}A^{r}+\partial_{i}A^{i}+\frac{r}{\ell^2}A_r=0,
\end{equation}
obtained by a $\nabla_{\nu}$ contraction of the equation (\ref{eqProca1g}).
\\
\\
Thus the following coupled field equations for $A_t$, $A_r$, and $A_i$, are obtained, 
\begin{eqnarray}\label{eq:P1g}
r^{\mp2z}\partial_{t}^{2}A_{t}\!-\!r^{2}\partial_{r}^{2}A_{t}\!-\!\partial_{i}\partial^{i}A_{t}\!-\!\left(\!1\!\mp 2z\!\right)r\partial_{r}A_{t}\mp 2zr\partial_{t}A_{r}\!+\!M^{2}\ell^{2}A_{t}=0,
\end{eqnarray}
\begin{eqnarray}\label{eq:P2g}
r^{\mp2z}\partial_{t}^{2}\!A_{r}\!-\!r^{2}\partial_{r}^{2}\!A_{r}\!-\!\partial_{i}\!\partial^{i}\!A_{r}\!-\!2r\partial_{r}\!A_{r}
\!-\!(1\!\pm\!2z)r^{\mp2z\!-\!1}\partial_{t}\!A_{t}\!+\!\frac{1\!+\!a_{i}}{r}\partial_{i}\!A{i}\!+\!M^{2}\ell^{2}\!A_{r}\!=\!0,
\end{eqnarray}
\begin{eqnarray}\label{eq:P3g}
r^{\mp2z}\partial_{t}^{2}A_{k}\!-\!r^{2}\partial_{r}^{2}A_{k}\!-\!\partial_{i}\partial^{i}A_{k}\!-(1\pm 2z)r\partial_{r}A_{k}\!\pm\!2zr\partial_{k}A_{r}\!+\!M^{2}\ell^{2}A_{k}\!=\!0,
\end{eqnarray}
where $A_t= A_t (t, r, x_ {i})$, $A_r= A_r (t, r, x_ {i}) $ and $A_k= A_k (t, r, x_ {i})$. If we further consider, for simplicity, that the only non-zero component of the vector potential is $A_t$ (following \cite{Taylor,LL}), then it follows from equation (\ref{eq:P2g}) that $\partial_{t}A_{t}=0$. Therefore, the field equation (\ref{eq:P1g}) for the $A_t$ component of the vector potential adopts the form  (see \cite{GHLMM,AHA} for a similar treatment of the perturbed vector field in a curved bulk background within the braneworld paradigm):
\begin{equation}\label{eq:unfourierg}
\left[ r^2\partial_r^2 + \!\left(\!1\!\mp2z\!\right)r\partial_r+ r^{-a_{i}}\partial_i^2-M^2\ell^2 \right] A_t(r,x^i) =0.
\end{equation}
Fourier transforming with respect to $x^i$, the equation (\ref{eq:unfourierg}) becomes an ordinary differential equation
\begin{equation}\label{eq:fourierg}
\left[r^2\partial_r^2 + \!\left(\!1\!\mp 2z\!\right)r\partial_r-k_{i}^{2}r^{-a_{i}}-M^2\ell^2\right]A_t(r,k_i) =0,
\end{equation}
which can be solved analytically for $D\ge 1$. The transformation 
\begin{equation}\label{eq:ansatzR}
A_{t}(r,k_{i})=r^{\frac{1}{2}(\pm 2z-1)}a_{t}(r,k_{i}),
\end{equation}
yields the following differential equation for the function $a_{t}(r,k_{i})$
\begin{equation}\label{eq:R1}
-\frac{\partial^{2}a_{t}(r,k_{i})}{\partial r^2} + \!\left(\frac{4k_{i}^{2}}{r^{a_{i}}}+4z^{2}-1+4M^{2}\ell^{2}\!\right)\frac{a_{t}(r,k_i)}{4r^2} =0,
\end{equation}
which has a  Schr$\ddot{o}$dinger-like form with the potential $V(r)=\frac{1}{4r^2}\left(\frac{4k_{i}^{2}}{r^{a_{i}}}+4z^{2}-1+4M^{2}\ell^{2}\right)$.
\vskip 4mm
Thus, for  $D=1$ the exact solution for this equation for arbitrary values of $z$ reads
\begin{equation}\label{eq:S1}
a_t(r) = r^{\frac{1}{2}}\left[c_{1}I_{\alpha}\left(\beta r^{-\frac{a_{1}}{2}}\right)+c_{2}I_{-\alpha}\left(\beta r^{-\frac{a_{1}}{2}}\right)\right] ,
\end{equation}
where, $c_1$ and $c_2$ are integration constants, $I_{\pm\alpha}\left(\beta r^{-\frac{a_{1}}{2}}\right)$ represent modified Bessel functions of first class and order $\pm\alpha$, with
\begin{eqnarray}
\alpha=\pm\frac{2\sqrt{M^2\ell^2+z^2}}{a_{1}}, \qquad
\beta=\frac{2k_{1}}{a_{1}} .
\end{eqnarray}
If we insert the function $a_{t}(r)$ into (\ref{eq:ansatzR}) we obtain for the vector field the following expression
\begin{equation}\label{eq:fourier4}
A_t(r) = r^{\pm z}\left[c_{1}I_{\alpha}\left(\beta r^{-\frac{a_{1}}{2}}\right)+c_{2}I_{-\alpha}\left(\beta r^{-\frac{a_{1}}{2}}\right)\right] .
\end{equation}
By expressing the modified Bessel functions of first class in terms of infinite series we have
\begin{eqnarray}
I_{\alpha}\left(\beta r^{-\frac{a_{1}}{2}}\right)= \sum_{s=0}^{\infty}\frac{1}{s!(s+\alpha)!}\left( \frac{\beta r^{-\frac{a_{1}}{2}}}{2}\right)^{2s+\alpha}, 
\end{eqnarray}
\begin{eqnarray}
I_{-\alpha}\left(\beta r^{-\frac{a_{1}}{2}}\right)= \sum_{s=0}^{\infty}\frac{1}{s!(s-\alpha)!}\left( \frac{\beta r^{-\frac{a_{1}}{2}}}{2}\right)^{2s-\alpha}, 
\end{eqnarray}
Furthermore, in order to have real solutions, we require the order of the Bessel functions to be real, hence the radicand of $\alpha$ must be positive leading to
\begin{equation}\label{eq:mbfg}
M^2\ge -\frac{z^2}{\ell^2}, 
\end{equation}
which is the Breitenlohner-Freedman bound for our Lifshitz spacetime when $D=1$.
\vskip 4mm
For the $D\ge 2$ cases we have performed a detailed, but not exhaustive, analysis of the Schr$\ddot{o}$dinger-like potential of equation (\ref{eq:R1}) for the allowed values of the $a_{i}$ constants, and have found a plethora of potentials that range from potential wells to potential barriers. Among them we have identified a family of potential wells with infinite walls that necessarily include negative energies and resemble the structure of the harmonic oscillator potential. Thus, the corresponding Schr$\ddot{o}$dinger-like equation for this class of potentials can be solved analytically and allow us to make a complete analysis of the respective mass spectrum for concrete values of the dynamical exponent $z$ and the constants $a_{i}$.
 
For the cases when $D=2,3,4$  we obtain the following Schr$\ddot{o}$dinger-like master equation  for the values of the $a_{i}$ and $z$ displayed in  Table 1:
\begin{equation}\label{eq:R2}
\!-\!\frac{\partial^{2}a_{t}(r, k_{i}^2)}{\partial r^2} \!+\! \!\left(\!\xi^{2}_{D,z}r^{2} \!+\!\frac{z^{2}\!-\!\frac{1}{4}\!+\!M^{2}\ell^{2}}{\!r^{2}}\!\right)a_{t}(r, k_{i}^2)=-\zeta^{2}_{D,z}a_{t}(r, k_{i}^2),
\end{equation}
where $\xi_{D,z}$ and $\zeta_{D,z}$ depend on $k_{i}^2$ according to Table 1. 
\begin{table}[H]
\centering
\begin{tabular}{ccccc}\hline
$D$  & \multirow{1}{1cm}{\centering $z$}& $a_{i}$ & $\xi_{D,z}\left(k_{i}^2\right)$ & $\zeta_{D,z}\left(k_{i}^2\right)$ \\
\hline
2 & 3 &$a_{1}\!=\!-2$,\quad  $a_{2}\!=\!-4$& $k_{2}^{2}$ & $k_{1}^{2}$ \\
\hline
\multirow{2}{1cm}{\centering 3} &4 &$a_{1}\!=\!-2$,\quad $a_{2}\!=\!-4$,\quad $a_3\!=\!-2$&$k_{2}^{2}$ &$k_{1}^{2}+k_{3}^{2}$  \\ 
 & 5& $a_{1}\!=\!-2$,\quad $a_{2}\!=\!-4$, \quad $a_3\!=\!-4$& $k_{2}^{2}+k_{3}^{2}$ & $k_{1}^{2}$ \\ 
\hline
\multirow{3}{1cm}{\centering 4} &5 &$a_{1}\!=\!-2$,\quad $a_{2}\!=\!-4$,\quad $a_3\!=\!-2$,\quad $a_4\!=\!-2$&$k_{2}^{2}$ &$k_{1}^{2}+k_{3}^{2}+k_{4}^{2}$  \\
 & 6 &$a_{1}\!=\!-2$,\quad $a_{2}\!=\!-4$,\quad $a_3\!=\!-2$,\quad $a_4\!=\!-4$& $k_{2}^{2}+k_{4}^{2}$ & $k_{1}^{2}+k_{3}^{2}$ \\
 & 7&$a_{1}\!=\!-2$,\quad $a_{2}\!=\!-4$,\quad $a_3\!=\!-4$,\quad $a_4\!=\!-4$& $k_{2}^{2}+k_{3}^{2}+k_{4}^{2}$ & $k_{1}^{2}$ \\ 
\hline
\end{tabular}\label{tab:Tabla1}
\caption{The special values of $D$, $z$ and $a_{i}$ corresponding to the expressions for $\xi_{D,z}\left(k_{i}^2\right)$ and $\zeta_{D,z}\left(k_{i}^2\right)$ for which the Schr$\ddot{o}$dinger-like equation (\ref{eq:R2}) renders analytical solutions with finite energy.}
\end{table}
As the dimensionality of spacetime $D$ increases the number of cases which obeys the Schr$\ddot{o}$dinger-like master equation  (\ref{eq:R2}) increases as well. The exact solution for this equation is
\begin{equation}\label{eq:R2S'}
a_{t}(r, k_{i}^2)=e^{-\frac{\xi_{D,z}r^{2}}{2}} r^{2\rho-\frac{1}{2}}\left[\hat{c}_{1}U(n,\rho,(\xi_{D,z}r^2))+\hat{c}_{2}L_{-n}^{2\rho-1}(\xi_{D,z}r^2)\right],
\end{equation}
where $\hat{c}_{1}$ and $\hat{c}_2$ are integration constants, $U(n,\rho,k_{2}r^2)$ is the confluent hypergeometric function and $L_{-n}^{2\rho-1}(k_{2}r^2)$ denote generalized Laguerre polynomials, with
\begin{eqnarray}
n=\frac{\zeta_{D,z}^2}{4\xi_{D,z}}+\rho,\qquad \rho=\frac{1}{2}(1+\sqrt{M^{2}\ell^{2}+z^2}).
\end{eqnarray}
However, it can be shown that the latter functions are divergent and therefore we shall set $\hat{c}_{2}=0$. 
Hence, we shall consider only the term containing the confluent hypergeometric function as the solution to the Schr$\ddot{o}$dinger-like equation (\ref{eq:R2}). \\
By substituting $a_{t}(r,\!\xi_{D,z}\!,\!\zeta_{D,z}\!)$ into (\ref{eq:ansatzR}), we get the final expression for the generic $A_{t}$ 
\begin{equation}\label{eq:R2S}
A_{t}(r, k_{i}^2)=\hat{c}_{1}e^{-\frac{\xi_{D,z}r^{2}}{2}} r^{\pm z+2\rho-1}U(n,\rho,(\xi_{D,z}r^2)).
\end{equation}
\\
\begin{figure}\centering\sidecaption \resizebox{0.85\hsize}{!}{\includegraphics*{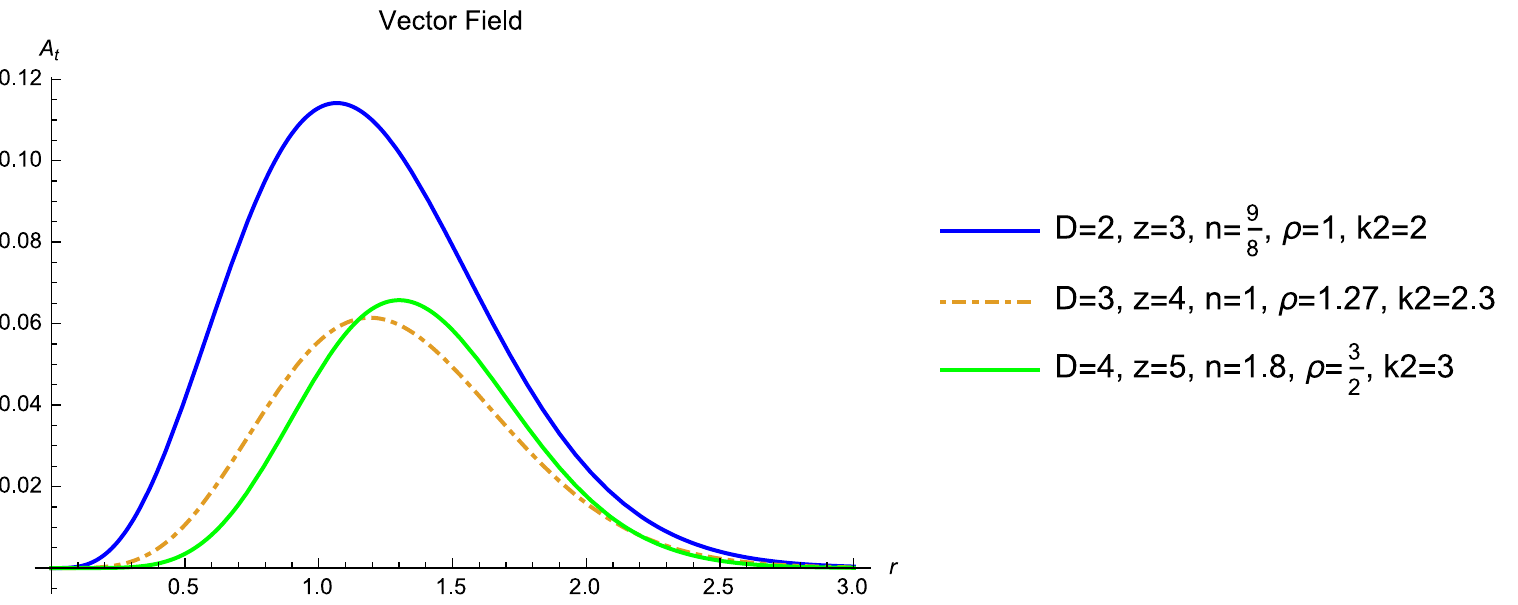}} \caption{\label{fig:i}The profiles of $A_{t}$ associated to $D=2$, $D=3$, $D=4$, where $\hat{c}_{1}=1$ for all cases.}
       \end{figure} 
 \\
It is remarkable that for these solutions to be real, the same bound (\ref{eq:mbfg}) must be obeyed for all the cases considered in Table 1. Thus we have obtained the same Breitenlohner-Freedman bound as for the $D=1$ case.\\
Here we would like to comment on an alternative approach to derivative the Breitenlohner-Freedman bound for our system.
In \cite{LL} the authors start from the Proca equations (\ref{eqProca1g}) and they consider only the electric component $A_t= A_{t}(r)\,dt$ on the bulk. By following them and using the metric given by (\ref{eqmg}), we find that
\begin{equation}\label{eq:Atmm}
A_t(r)=\check{c}_1 r^{\tau_{\pm}+\lambda_{+}}+ \check{c}_2 r^{\tau_{\pm}+\lambda_{-}},
\end{equation}
where $\tau_{\pm}=\pm z$ and $\lambda_{\pm}=\pm \sqrt{M^2\ell^{2}+z^2}$. 
Thus, we have obtained one more time the same Breitenlohner-Freedman bound (\ref{eq:mbfg}) for the massive vector field. 

Therefore we can conclude that the mass spectrum of the  solution given by (\ref{eqm2g}) satisfies the Breitenlohner-Freedman bound upon substitution into (\ref{eq:mbfg}) in a straightforward way and it holds for any dimension $D$ when $A_{t}=A_{t}(r,k_i)$ and in this sense yields a stable field configuration.

By following the same argument as for the AdS/CFT correspondence in \cite{Witten}, one can interpret $\check{c}_{2}$ in (\ref{eq:Atmm}) as the normalisable mode for dual operator of Lifshitz dimension $\tau_{-}+\lambda_{-}$.
\vskip 4mm

\subsection{The null energy condition (NEC)}
Einstein's equations can be obtained without specifying matter, such that the $ T _ {\mu\nu} $ is arbitrary. Thus, understanding properties of Einstein's equations that have a certain variety of sources (fields) is very important. Therefore, it is convenient to impose energy conditions that restrict the arbitrariness of $ T _ {\mu\nu} $. Hence, to calculate these energy conditions we must construct scalars of $ T _ {\mu\nu} $, by contracting it with arbitrary timelike vectors $ \zeta^{\mu} $ or arbitrary null vectors $ \xi^{\mu} $.

It is important to mention that there are several types of energy conditions, which are appropriate for different circumstances. In our case the NEC is generally assumed to provide a sufficient condition to have a physically sensible holographic dual in the semi-classical limit \cite{Robert}. 

Although one can calculate the NEC using the energy-momentum tensor, one can also write this condition in terms of the Ricci tensor. Following \cite{FosterLiu}, the NEC in terms of the Ricci tensor $R_{\mu\nu}$ and an arbitrary null vector, is as follows
\begin{equation}\label{eq:null}
R_{\mu\nu}\xi^{\mu}\xi^{\nu}\ge 0,
\end{equation}
or
\begin{eqnarray}\label{eq:null1}
R_{tt}(\xi^{t})^2+R_{rr}(\xi^{r})^2+R_{ij}\xi^{i}\xi^{j}\ge 0.
\end{eqnarray}
If we use our background given by (\ref{eqmg}), a straightforward computation of the components of the Ricci tensor gives 
\begin{eqnarray}\label{Ricci}
R_{tt}=-\frac{g_{tt}}{\ell^{2}}\left[\!z^{2}\!\pm\!\frac{z}{2}\displaystyle\sum_{k=1}^{D}a_{k}\!\right],\nonumber\\
R_{rr}=-\frac{g_{rr}}{\ell^{2}}\left[\!z^{2}\!+\!\frac{1}{4}\displaystyle\sum_{k=1}^{D}a_{k}^{2}\!\right],\\
R_{ij}=-\frac{g_{ij}}{\ell^{2}}\left[\!\pm z\!+\!\frac{1}{2}\displaystyle\sum_{k=1}^{D}a_{k}\!\right]\frac{a_{i}}{2}.\nonumber
\end{eqnarray}
By substituting these expressions into (\ref{eq:null1}) we obtain
\begin{eqnarray}\label{eq:null22}
\left[\!z^{2}\!\pm\!\frac{z}{2}\displaystyle\sum_{k=1}^{D}a_{k}\!\right]r^{\pm 2z}(\xi^{t})^2\!-\!\left[\!z^{2}\!+\!\frac{1}{4}\displaystyle\sum_{k=1}^{D}a_{k}^{2}\!\right]r^{\!-\!2}(\xi^{r})^2\!-\!\left[\!\pm z\!+\!\frac{1}{2}\displaystyle\sum_{k=1}^{D}a_{k}\!\right]\frac{a_{i}}{2}r^{a_{i}}(\xi^{i})^2\ge 0.
\end{eqnarray}
Subsequently, by substituting the explicit expressions of $\displaystyle\sum_{k=1}^{D}a_{k}$ and $\displaystyle\sum_{k=1}^{D}a_{k}^{2}$, which are given by (\ref{eq:sum}) and (\ref{eq:ac}), respectively, into the inequality (\ref{eq:null22}), we get
\begin{equation}\label{eq:null3}
-\!D(D+1)\ell^{2}r^{\!-\!2}(\xi^{r})^2\ge 0.
\end{equation}

Thus, the inequality (\ref{eq:null3}) can not be satisfied for an arbitrary null vector since it vanishes only when $\xi^{r}$ is identically zero. In other words, the NEC is only satisfied when applied along the $t-x_{i}$ directions. It is important to emphasize that even when the NEC is not satisfied for all directions of the bulk spacetime i.e. for $t, r, x_{i},$ where $i=1,2,3, ..., D$, it does along the $ D + 1 $ directions of the boundary, i.e. in the time direction $ t $ and the spatial directions $x_{i}$ of the field theory.


\subsection{Spatially isotropic Lifshitz background}

In Section 3, we showed the general construction of a new family of spatially anisotropic Lifshitz spacetimes with arbitrary dynamical exponent $z$ and constant negative scalar curvature in d+1 dimensions within the framework of the Einstein-Proca theory. 

Nonetheless, it is possible to obtain a spatially isotropic Lifshitz background from the results obtained in the previous Section.  By construction, starting from the same action of the Section 3 and making $P_{i}(r)=p(r)$, we get the following metric anzatz
\begin{equation}
\label{eq:metrica}
ds^{2} = \ell^{2} \left( -f(r) dt^{2} + p(r) dx^{2}_{i} + \frac{dr^{2}}{r^{2}}\right),
\end{equation}
where again $i = 1, 2, ..., D$. If we further consider
\begin{equation}\label{eq:f} 
f(r) = r^{\pm2z},
\end{equation}
just as in our previous analysis, as a result of substituting these \textit{ans$\ddot{a}$tze} into the Einstein-Proca field equations (\ref{eq tt 2g})-(\ref{Procaeqng}), we obtain the following Lifshitz solutions:
\begin{equation} \label{eq:p}
p(r) =  r^{\mp\frac{(D-1)z}{D^2}}
\end{equation}
\begin{equation} \label{ka}
A_t= c\, r^{\pm z}, 
\end{equation}
where the power of $p(r)$ shows a dependence on the spacetime dimensionality $D$ and the critical exponent $z$,  and the non-vanishing component of the massive vector field $A_{t}$ has an arbitrary real constant $c$. In addition, the parameters of the theory read
\begin{equation} \label{eq:m2}
M^2 = - \frac{2 D^2 (D-1)}{3D-1},
\end{equation} 

\begin{equation} \label{k}
c = \sqrt{ \frac{(3D-1)(2D-1)(D+1)}{4D^5}}\,z,
\end{equation}

\begin{equation}\label{eq:l}
\ell^2 = \frac{(3D-1)z^2}{4D^3}.
\end{equation}

It is important to emphasize, that this Lifshitz solution possesses a spectrum with negative squared masses as in the spatially anisotropic case. As pointed out before, these modes of the massive vector field must satisfy a Breitenlohner-Freedman bound. This issue will be discussed in the next subsection.
\\
\\
Thereby, the full metric is expressed as follows
\begin{equation}\label{eq:m}
ds^{2}=\ell^2\left(-r^{\pm2z}dt^{2}+\frac{1}{r^{2}}dr^{2}+r^{\mp\frac{(D-1)z}{D^{2}}} dx^{i}dx_{i}\right),
\end{equation}
which is invariant under the Lifshitz transformations (\ref{eq3}) if the coordinates transform in the following way:
 \begin{eqnarray}
& & r \longrightarrow r'=\lambda^{\pm} r, \\
&  & t \longrightarrow t'=\lambda^{\mp z}t,  \\
&  & x^{i}\longrightarrow x^{'i}=\lambda^{\pm{\frac{(D-1)z}{2D^2}}} x^{i} \, \label{eq:ansatz_e}.
\end{eqnarray}
\vskip 3mm
The curvature scalar upon substitution of the metric functions $f$ and $p$, and the expression for the Lifshitz curvature radius $\ell$ in (\ref{eqRg}), reads
\begin{equation}
R=-\frac{\left( D+1\right) \left( 5D^2-2D+1\right) }{\left( 3D-1\right) };
\end{equation}
here we note that the curvature scalar only depends on the dimensionality $D$, it is constant and negative definite.
\\

It is worth mentioning that if $t$, $r$ and $x_{i}$ transform in the following way:
\begin{eqnarray}\label{re}
r&=&\tilde{r}^{-\frac{2D^2}{(D-1)z}},\nonumber\\
t&=&\tilde{t},\\
x_{i}&=&\tilde{x_{i}},\nonumber
\end{eqnarray}
and $z=\frac{-2D^2}{D-1}$, the metric (\ref{eq:m}) reads

\begin{equation}\label{m2}
ds^{2}=\ell^2\left(-\tilde{r}^{\pm2z}d\tilde{t}^{2}+\frac{d\tilde{r}^{2}}{\tilde{r}^{2}}+\tilde{r}^{\pm 2}d\tilde{x}_{i}^{2}\right),
\end{equation}
concluding that the spatially isotropic Lifshitz background given by (\ref{eq:m}) is diffeomorphic to the one reported in \cite{Taylor}.
\subsubsection{BF bound for our spatially isotropic Lifshitz background}

To find the Breitenlohner-Freedman bound, we consider a spacetime given by the background (\ref{eq:m}) and the Proca equations without gravitational back-reaction, i.e.,  in the perturbative limit in which the massive vector field does not alter the structure of spacetime. Thus, the Proca equations read
\begin{equation}\label{eq:Proca1}
\nabla_\mu F^{\mu \nu} - M^2 A^\nu = 0 \,,
\end{equation}
and are supplemented by the following constraint  in curved spacetime
\begin{equation}\label{eq:lorentz}
\nabla_{\nu}A^{\nu}=\partial_{0}A^{0}+\partial_{r}A^{r}+\partial_{i}A^{i}+\left[1\pm\frac{(D+1)z}{2D}\right]\frac{r}{\ell^2}A_r=0,
\end{equation}
obtained by a $\nabla_{\nu}$ contraction of the equation (\ref{eq:Proca1}).
\\
Hence the following coupled field equations for $A_r$, $A_t$, and $A_i$, are obtained, 
\begin{eqnarray}\label{eq:P1}
r^{\mp2z}\partial_{t}^{2}A_{t}\!-\!r^{2}\partial_{r}^{2}A_{t}\!-\!r^{\pm{\frac{z\!(D\!-\!1\!)}{D^2}}}\partial_{i}^{2}\!A_{t}\!-\!\left(\!1\!\mp\!\frac{(\!3D\!-\!1\!)z}{2D}\!\right)r\partial_{r}\!A_{t}\mp 2zr\partial_{t}A_{r}\!+\!M^{2}\!\ell^{2}\!A_{t}=0,
\end{eqnarray}
\begin{eqnarray}\label{eq:P2}
r^{\mp2z}\partial_{t}^{2}A_{r}-r^{2}\partial_{r}^{2}A_{r}\!-\!r^{\pm{\frac{z(D-1)}{D^2}}}\partial_{i}^{2}A_{r}\!-\!\left(\!2\pm\frac{(D+1)z}{2D}\!\right)r\partial_{r}A_{r}-\\ \nonumber
(1\pm 2z)r^{\mp2z-1}\partial_{t}A_{t}+\left(1\mp\frac{(D-1)z}{D^2}\right)r^{-1}\partial_{i}A^{i}+M^{2}\ell^{2}A_{r}=0,
\end{eqnarray}
\begin{eqnarray}\label{eq:P3}
r^{\mp2z}\partial_{t}^{2}A_{k}-r^{2}\partial_{r}^{2}A_{k}-r^{\pm{\frac{z(D-1)}{D^2}}}\partial_{i}^{2}A_{k}-\left(1\pm\frac{(2D-1)(D+1)z}{2D^2} \right)r\partial_{r}A_{k}\pm \\ \nonumber
\frac{(D^{2}-1)z}{2D^2}r\partial_{k}A_{r}+M^{2}\ell^{2}A_{k}=0,
\end{eqnarray}
where $A_t= A_t (t, r, x_ {i})$, $A_r= A_r (t, r, x_ {i}) $ and $A_k= A_k (t, r, x_ {i})$. If we further consider, for simplicity, that the only non-zero component of the vector potential is $A_t$, then it follows from equation (\ref{eq:P2}) that $\partial_{t}A_{t}=0$. Therefore, the field equation (\ref{eq:P1}) for the component $A_t$ of the vector potential adopts the form:
\begin{equation}\label{eq:unfourier}
\left\lbrace  \partial_r^2 + \left[ 1\mp\frac{(3D-1)z}{2D}\right] r^{-1}\partial_r+ r^{\pm{\frac{(D-1)z}{D^2}-2}}\partial_i^2-\ell^2 M^2 r^{-2}\right\rbrace A_t(r,x^i) =0.
\end{equation}
Fourier transforming with respect to $x^i$, the equation (\ref{eq:unfourier}) becomes an ordinary differential equation
\begin{equation}\label{eq:fourier}
\left\lbrace  \partial_r^2 + \left[ 1\mp\frac{(3D-1)z}{2D}\right] r^{-1}\partial_r- k^2 r^{\pm{\frac{(D-1)z}{D^2}}-2}-\ell^2 M^2 r^{-2}\right\rbrace A_t(r,k_i) =0,
\end{equation}
where $k^2=k_{i}k^{i}$ and the exact solution for this equation for general values of $z$ is
\begin{equation}\label{eq:fourier4}
A_t(r) = r^{\pm\frac{(3D-1)z}{4D}} \left[  c_{+} I_{\gamma}\left( \eta r^{\pm\frac{(D-1)z}{2D^2}}\right)  + c_{-} I_{-\gamma}\left( \eta r^{\pm\frac{(D-1)z}{2D^2}}\right) \right] ,
\end{equation}
where, $c_{+}$ and $c_{-}$ are integration constants, $I_{\pm\gamma}\left( \eta r^{\pm\frac{(D-1)z}{2D^2}}\right)$ represent modified Bessel functions of first class and order $\pm\gamma$, with 
\begin{eqnarray}
\gamma&=&\pm \frac{D\sqrt{16D^2 \ell^2 M^2+ (3D-1)^2 z^2}}{2(D-1)z}, \\
\eta&=& \pm\frac{2D^2k}{(D-1)z}.
\end{eqnarray}

Furthermore, in order to have real solutions, we require the  order of the Bessel functions to be real, hence  
\begin{equation}
16D^2 \ell^2 M^2+ (3D-1)^2 z^2\ge 0,
\end{equation}
by substituting the value of $ \ell ^ 2 $ given by (\ref {eq:l}) into this expression, we get
\begin{equation}\label{eq:mbf}
M^2\ge -\frac{D (3D-1)}{4}, 
\end{equation}
which is the Breitenlohner-Freedman bound for our spatially isotropic Lifshitz spacetime.
\vskip 2mm
An alternative way of obtaining this result is as follows:
\\
If we start from the Proca equations (\ref{eq:Proca1}), considering only the electric component $A= A_{t}(r)\,dt$ on the bulk and using the metric given by (\ref{eq:m}), we find that
\begin{equation}
A_t(r)=\tilde{c}_1 r^{\kappa_{\pm}+\omega_{-}}+ \tilde{c}_2 r^{\kappa_{\pm}+\omega_{+}},
\end{equation}
where $\kappa_{\pm}=\pm\frac{(3D-1)z}{4D}$ and $\omega_{\mp}=\mp\frac{\sqrt{16D^2\ell^2 M^2+(3D-1)^2 z^2}}{4D}$. 
\\
By substituting the value of $\ell^2$ (\ref {eq:l}) into the radicand appearing in $\omega_{\mp}$, we get the same Breitenlohner-Freedman bound (\ref{eq:mbf}) for the massive vector field. Besides, one can interpret $\tilde{c}_{1}$ and $\tilde{c}_{2}$ as the normalisable modes for dual operators of Lifshitz dimension $\kappa_{-}+\omega_{\mp}$.
\\

The mass spectrum for this solution given by (\ref{eq:m2}) satisfies the BF bound upon substitution into (\ref{eq:mbf}) in a straightforward way, rendering an inequality, $(D+1)^2\geq 0$, which holds for any dimension $D$ and yields a stable field configuration.
\vskip 2mm

\subsubsection{Null energy condition (NEC)}
In terms of the Ricci tensor $R_{\mu\nu}$ and an arbitrary null vector, the NEC reads
\begin{equation}\label{eq:nulli}
R_{\mu\nu}\xi^{\mu}\xi^{\nu}\ge 0.
\end{equation}
For the spatially isotropic case, the Ricci tensor components look like
\begin{eqnarray}\label{Riccii}
R_{tt}=-\left(\frac{D+1}{2D}\right)z^{2}\frac{g_{tt}}{\ell^{2}},\nonumber\\
R_{rr}=-\left(1+\frac{(D-1)^2}{4D^3} \right)z^2\frac{g_{rr}}{\ell^{2}},\\
R_{ij}=\left(\frac{D^2-1}{4D^3}\right)z^2\frac{g_{ij}}{\ell^{2}}.\nonumber
\end{eqnarray}
By substituting these Ricci tensor components into (\ref{eq:nulli}) we obtain
\begin{eqnarray}\label{eq:null2}
z^2\frac{(D\!+\!1)}{2D}r^{\pm 2z}(\xi^{t})^2\!-\!z^2\left(1\!+\!\frac{(D\!-\!1)^2}{4D^3} \right)r^{\!-\!2}(\xi^{r})^2\!+\!z^2\left(\frac{D^2\!-\!1}{4D^3}\right)r^{\mp \frac{(D\!-\!1)z}{d^2}}(\xi^{i})^2\ge 0.
\end{eqnarray}
As in subsection 3.3, for this particular spatially isotropic Lifshitz spacetime, the NEC can not be satisfied for an arbitrary null vector since it vanishes only when $\xi^{r}$ is identically zero. In other words, the NEC is only satisfied when applied along the boundary directions, i.e. along time direction $ t $ and the $x_{i}$ spatial directions of the field theory. 

\subsubsection{Another look at spatially isotropic Lifshitz backgrounds}
Now, we show that if one considers $P_{i}=r^{2}$, the metric ansatz is expressed as follows
\begin{equation}\label{m23}
ds^{2}=\ell^2\left(-f(r)dt^{2}+\frac{dr^{2}}{r^{2}}+r^{2}dx_{i}^{2}\right).
\end{equation}
As a result of substituting the function $P_i=r^{2}$ into the Einstein-Proca field equations (\ref{eq tt 2g})-(\ref{Procaeqng}), we obtain the following function $f(r)$ as a solution
\begin{eqnarray}\label{fk}
f(r)=r^{2z}, 
\end{eqnarray}
where the parameter $z$ is the critical exponent and can be positive or negative in principle. By substituting (\ref{fk}) in (\ref{m23}), we obtain
\begin{equation}\label{eq:mdk}
ds^{2}=\ell^2\left(-r^{2z}dt^{2}+\frac{1}{r^{2}}dr^{2}+r^{2} dx^{i}dx_{i}\right).
\end{equation}
where the parameters of the theory read
\begin{eqnarray} \label{eq:lk}
\ell^2 = \frac{(2z+D-1)^{2}+(3D-1)(D+1)}{4D(D+1)},
\end{eqnarray} 
\begin{eqnarray} \label{eq:mk}
M^2 = \frac{2 D^2 (D+1)}{(2z+D-1)^{2}+(3D-1)(D+1)}z, 
\end{eqnarray} 
\begin{eqnarray} \label{eq:ck}
c= \sqrt{ \frac{2D(z-1)}{M^{2}}},
\end{eqnarray} 
and the massive vector field is expressed by
\begin{equation}\label{ak}
A_{t}=cr^{z}.
\end{equation}
As one can see, the massive vectorial field is real if $z>1$, which implies that $M^2$ is positive too. It is convenient to stress that at this point, the spatially isotropic Lifshitz background reported in \cite{Taylor} is recovered. Whereas if $z\leq 0$ the massive vectorial field turns out be real if $M^2$ is negative, since $M^2$ is proportional to $z$. For $0<z<1$ there is no real vector field solution.
\\

Finally, the curvature scalar upon substitution of the metric functions $f$ and $p$ in (\ref{eqRg}), reads
\begin{equation}
R=-\frac{2D(D+1)\left[4z^{2}+2D(2z+D+1)\right]}{(2z+D-1)^{2}+(3D-1)(D+1)};
\end{equation}
here we note that the curvature scalar depends on the dimensionality $D$ and the critical exponent $z$, it is negative and constant for any $z$.

\section{Conclusions and discussion}
We have obtained a new family of exact solutions to the Einstein-Proca equations with a \textit{single} vector field that have spatially anisotropic Lifshitz symmetry, for any dimension $D$ and arbitrary dynamical exponent $z$. Our functions $P_{i}(r)$, which multiply $dx_{i}^{2}$ in the metric, depend on the dimension $D$, the curvature Lifshitz radius $\ell^2$ and the critical exponent $z$, giving rise to a spatially anisotropic generalization of the known Lifshitz backgrounds constructed within the Einstein-Proca theory, since in the spacetime metric all the $x^{i}$ coordinates scale in a different way. Thus we have obtained an interesting new family of spatially anisotropic Lifshitz solutions that is different from those supported by several vector fields \cite{Taylor1}. In addition, we have found that the squared mass of the Proca field turns out to be negative definite, making it necessary to obtain a generalized Breitenlohner-Freedman bound for our theory. We have computed such a bound  in two different ways (one of them considers a full dependence of the vector field $A_t$ on all the spacetime coordinates) and have shown that our mass spectrum respects it, yielding a stable and physically meaningful field configuration.
We found that our curvature scalar is also negative definite, depends only on the dimensionality $D$ and is completely independent of the dynamical exponent $z$. 
It would be interesting to obtain further generalizations of these Lifshitz backgrounds where the curvature scalars depending on both the dimensionality $D$ and the critical exponent $z$. 

On the other hand, we also obtained a spatially isotropic Lifshitz background, given by (\ref{eq:m}), built within the Einstein-Proca theory supported by a \textit{single} vector field. This spatially isotropic spacetime is diffeomorphic to a Lifshitz geometry with negative $z$ as shown in Section $3.3$. 

We further analyzed the null energy condition of both the spatially anisotropic Lifshitz spacetime (\ref{eqmg}) and the spatially isotropic Lifshitz background (\ref{eq:m}). It is worth mentioning this condition since it plays a relevant role in the sense that we can obtain restrictions for the dynamical exponents determined by the NEC and these restrictions will be applied when studying the invariant fixed points in the field theories. The NEC is remarkable because it plays a crucial role when examining the behavior of the flow of the c-function, since that the NEC must be satisfied for the flow of the c-function to be monotonic for dual field theories to Einstein gravity, for more details see \cite{Robert}.

Regarding the spatially anisotropic Lifshitz geometry (\ref{eqmg}), the NEC is satisfied for all the boundary directions, i.e. along the time direction $ t $ and the spatial directions $x_{i}$ of the dual field theory with the corresponding conditions on the scaling exponents. However, the NEC is not satisfied for a completely arbitrary null vector due to the presence of negative critical exponents.
Since the spatially isotropic Lifshitz spacetime (\ref{eq:m}) possesses a negative dynamical exponent it also violates the NEC. However, in \cite{hoyos} the authors presented Lifshitz backgrounds obtained within a gravitational model with curvature squared corrections where solutions with $z<1$ also are allowed and, in principle, the NEC can be satisfied for negative dynamical exponents. This implies that so far there is no exhaustive study about whether or not the NEC holds for Lifshitz solutions with $z<1$, opening a window for gravitational configurations for which the NEC can be satisfied. It would be interesting to either confirm or refute this hypothesis within the context of our model, i.e. by adding curvature squared corrections to the Einstein-Proca theory and see if we can obtain Lifshitz spacetimes where the null energy condition can be satisfied for negative critical exponents. We leave this issue for future work.

One of the most important results in this spatially  isotropic case is the negative definite character of the squared mass of the Proca field which supplements the positive definite squared mass spectra known so far in the literature. We also have found a Breitenlohner-Freedman bound for this background and shown that the mass spectrum respects it, yielding a stable field configuration. Likewise, its curvature scalar was calculated, it is also negative definite and depends only on the dimensionality $D$. A detailed study of the physical consequences of the negative squared mass spectrum on the dual quantum field theory is in order here to better understand this apparent inconsistency of the aforementioned spatially isotropic family of Lifshitz spaces.

Finally, in order to be able to apply Lifshitz holography to condensed matter systems one needs to understand in detail the correspondence existing between bulk fields and boundary operators, i.e. to establish a holographic dictionary.  However, this dictionary is very subtle and is not completely well understood for non-relativistic theories as the Lifshitz one. Recent progress on this issue show that holographic models describe universality classes of Lifshitz theory at strong coupling and the holographic dictionary was used to deduce universal properties of certain Lifshitz systems \cite{Taylor}. It would be interesting to determine whether the negative character of the squared mass spectrum of our solution introduce some subtleties within a given dual field theory. This is a work currently in progress.


\section*{Acknowledgments}
 
The work of AHA was completed at the Aspen Center for Physics, which is supported by National Science Foundation grant PHY-1607611 and a Simons Foundation grant as well. He expresses his gratitude to the ACP for providing an inspiring and encouraging atmosphere for conducting part of this research. RCF and AHA acknowledge a VIEP-BUAP grant. RCF, AHA and JMR thank SNI for support. VMZ acknowledges a CONACYT PhD fellowship. UN acknowledges a fellowship granted by PROMEP-SEP and is grateful to SNI-CONACYT for a research assistant grant. RCF, AHA, VMZ and UN acknowledgments support by a CONACYT grant No. A1-S-38041.


%
%
%
%

\end{document}